\documentclass[10pt,twocolumn,superscriptaddress,preprintnumbers,amsmath,amssymb,prb,aps]{revtex4-1}
\usepackage{graphicx}
\usepackage{pgf}
\usepackage{ulem}
\usepackage{color}
\usepackage{hyperref}

\newcommand{\ITC}[1]{{\color{black} #1}}
\newcommand{\ITD}[1]{{\color{black} #1}}

\begin{document}

\title{
Coexistence of type-I and type-II superconductivity signatures in ZrB$_{12}$ probed by muon spin rotation measurements
} 

\author{P.~K.~Biswas}
\email[]{pabitra.biswas@stfc.ac.uk}
\affiliation{ISIS Pulsed Neutron and Muon Source, STFC Rutherford Appleton Laboratory, Harwell Campus, Didcot, Oxfordshire, OX11 0QX, United Kingdom}

\author{F.~N.~Rybakov}
\affiliation{Department of Physics, KTH Royal Institute of Technology, SE-10691 Stockholm, Sweden}

\author{R.~P.~Singh}
\affiliation{Indian Institute of Science Education and Research Bhopal, Bhopal, 462066, India}

\author{Saumya~Mukherjee}
\affiliation{Diamond Light Source, Oxfordshire OX11 0DE, United Kingdom.}
\affiliation{Clarendon Laboratory, Department of Physics, University of Oxford, Parks Road, Oxford OX1 3PU, United Kingdom}

\author{N.~Parzyk}
\affiliation{Physics Department, University of Warwick, Coventry, CV4 7AL, United Kingdom}

\author{G.~Balakrishnan}
\affiliation{Physics Department, University of Warwick, Coventry, CV4 7AL, United Kingdom}

\author{M.~R.~Lees}
\affiliation{Physics Department, University of Warwick, Coventry, CV4 7AL, United Kingdom}

\author{C.~D.~Dewhurst}
\affiliation{Institut Laue Langevin, 6 Rue Jules Horowitz, 38042 Grenoble, France}

\author{E.~Babaev}
\email[]{babaev.egor@gmail.com}
\affiliation{Department of Physics, KTH Royal Institute of Technology, SE-10691 Stockholm, Sweden}

\author{A.~D.~Hillier}
\affiliation{ISIS Pulsed Neutron and Muon Source, STFC Rutherford Appleton Laboratory, Harwell Campus, Didcot, Oxfordshire, OX11 0QX, United Kingdom}

\author{D.~M$^c$K.~Paul}
\affiliation{Physics Department, University of Warwick, Coventry, CV4 7AL, United Kingdom}


\begin{abstract}
Superconductors usually display either type-I or type-II superconductivity and the coexistence of these two types in the same material, \ITD{for example at different temperatures} is rare in nature. We the employed muon spin rotation ($\mu$SR) technique to unveil the superconducting phase diagram of the dodecaboride ZrB$_{12}$ and obtained clear evidence of both type-I and type-II characteristics.
Most importantly, we found a region showing unusual behavior where the usually mutually exclusive  $\mu$SR signatures of type-I and type-II superconductivity coexist. We reproduced that behavior in theoretical modeling that required taking into account multiple bands and multiple coherence lengths, \ITD{which suggests that material has one coherence length larger and another smaller than the magnetic field penetration length (the type-1.5 regime)}. 
\ITD{At stronger fields, a} footprint of the type-II mixed state showing square flux-line lattice was also obtained using neutron diffraction. 
\end{abstract}

\maketitle

\section{Introduction}

\ITD{Despite a century of research, the classification of superconducting states according to types of  fermionc pairing and resulting properties is an actively developing field.}

According to Ginzburg and Landau paradigm, superconductors are classified as type-I and type-II based on the distribution of the magnetic field in materials. This is defined by Ginzburg–Landau parameter, $\kappa=\frac{\lambda}{\xi}$, where $\lambda$ is the magnetic penetration depth and $\xi$ is the coherence length \cite{Tinkham}. Superconductors with $\kappa<\frac{1}{\sqrt{2}}$ are considered type-I which show expulsion of the magnetic field below $H_{\rm c}$, known as the Meissner state. More commonly, they exhibit non-zero demagnetization factor which energetically favors the coexistence of flux free superconducting regions (${B = 0}$) and the normal regions with finite internal field ${\approx\!B_{\rm c}}$. Such complex field texture is the representative feature of Type-I superconductors and is identified as the intermediate state described by Landau \cite{Landau}. For type-II superconductors where $\kappa>\frac{1}{\sqrt{2}}$, it is energetically favorable for the magnetic flux to penetrate the sample in the form of flux-lines parallel to the applied field also known as Abrikosov vortices. The repulsive interactions between these vortices tend to produce a stable lattice structure. This state is called a mixed state and is the main feature of a type-II superconductor \cite{Roulin}. The behavior is more complicated at the boundary condition of $\kappa=\frac{1}{\sqrt{2}}$ where, according to the Ginzburg-Landau model, the interactions between the vortex lines disappear at all distances~\cite{Kramer71,Bogom}. 
Microscopic corrections beyond Ginzburg-Landau lead to weak vortex clustering in the vicinity of $\kappa \approx 1/\sqrt{2}$ which is described as the  intermediate-mixed state (IMS) \cite{Jacobs73, Eilenberger69, Krageloh, Klein,Christen,Laver}. 
This mechanism of vortex attraction requires fine-tuning, namely the near-cancellation of density-mediated attractive and current-mediated repulsive intervortex forces, which is only possible in idealized single-band models~\cite{Jacobs73, Eilenberger69, Krageloh, Klein}. 
However, many superconductors are multiband or break multiple symmetries and thus have multiple coherence lengths. In such systems, the penetration depth can fall between two coherence lengths and aspects of coexistent type-I-like and type-II-like behavior originating from multiple components can arise~\cite{Babaev2,Silaev1}. In this regime, termed as type-1.5 \cite{Moshchalkov}, vortices are thermodynamically stable like in type-II superconductors, but have cores extending beyond the flux carrying area like in type-I superconductors, which leads to long-range attractive and short-range repulsive interactions~\cite{Babaev2,Babaev1,Silaev1,Silaev2,PhysicaC,winyard1}. 
Observation of this state was pursued experimentally \cite{Moshchalkov,Ray}, but its properties remain  little explored.

\begin{figure*}[ht]
\begin{center}
\includegraphics[width=2.0\columnwidth]{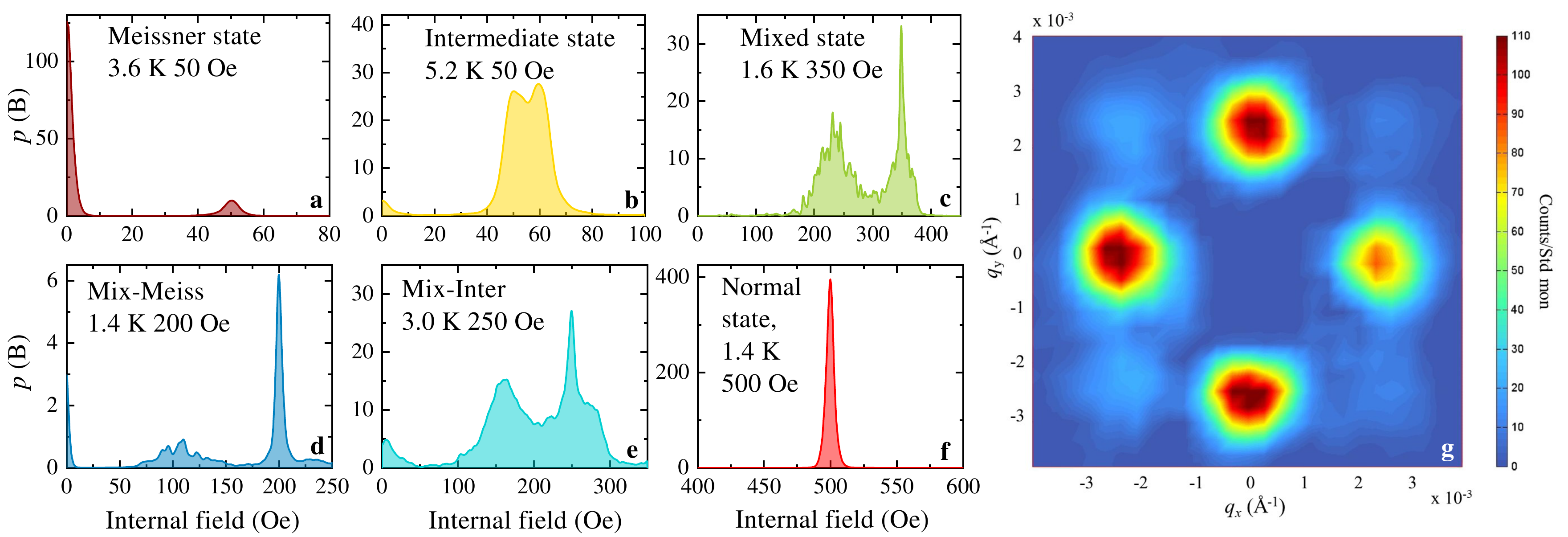}
\caption{Field distribution, $p(B)$, of the local field probed by muons illustrates the typical signal observed in the \textbf{a} Meissner, \textbf{b} intermediate, \textbf{c} mixed, \textbf{d} mixed-Meissner, \textbf{e} mixed-intermediate, and \textbf{f} normal states. The TF-$\mu$SR time spectra for the corresponding states are shown in the Supplemental Material. \textbf{g} Neutron diffraction image showing formation of a regular square flux-line lattice in ZrB$_{12}$, grown under an applied field of 300~Oe at 1.6 K.}
 \label{fig:FFT}
\end{center}
\end{figure*}

Superconductivity in ZrB$_{12}$ was first observed as early as in 1960's by Matthias~\textit{et al.}~\cite{Matthias, Chu}.
ZrB$_{12}$ has a relatively high $T_{\rm c}$ of 6.1~K within the family of the dodecaborides $Me$B$_{12}$ ($Me$ = Sc, Y, Zr, La, Lu, Th). Uniquely, ZrB$_{12}$ displays a variety of deviations from the conventional superconducting behavior. Several models have been proposed to explain its superconducting properties, including describing ZrB$_{12}$ as a conventional superconductor with enhanced surface effects~\cite{Tsindlekht, Khasanov}, a strongly coupled \textit{s}-wave gap structure~\cite{Daghero}.
Even \textit{d}-wave superconductivity has been proposed for this compound~\cite{Gasparov1}. However, the true nature of the superconductivity remains debatable and calls for more detailed study~\cite{Sluchanko}. The temperature dependence of magnetic penetration depth exhibits pronounced a two-band feature~\cite{Gasparov} which requires weak interband coupling~ \cite{twoband}. 
\ITD{This is precisely the condition where the system develops two well-defined coherence lengths, requires retaining two field components in a Ginzburg-Landau description~\cite{Silaev1,Silaev2}, and excludes the physics that relies on near degeneracy of the length scales~\cite{Jacobs73, Eilenberger69, Krageloh, Klein}, which make it a particularly interesting material to explore new physics.}
Recent band-structure calculations of ZrB$_{12}$ concluded that the Fermi surface of ZrB$_{12}$ is composed of an open sheet and a closed sheet~\cite{Gasparov2} which might be the source of the multi band character of superconductivity.
Furthermore, while, it was reported that specific heat measurements as a function of field show a pronounced first-order-like normal- to superconducting-state transition with latent heat at elevated temperatures, consistent with a type-I behavior~\cite{Wang}, different studies report observation of vortices at relatively high temperatures \cite{Junyi}. These contradicting results indicate ZrB$_{12}$ may represent a special class of superconductors, which are inherently capable of hosting the type-I and type-II \ITD{characteristics} simultaneously.
This has attracted our attention for investigating the superconducting properties of ZrB$_{12}$ using muon spin rotation/relaxation ($\mu$SR) spectroscopy. This technique has been widely used to map the phase diagram and study the microscopic properties of the vortex and Meissner state of superconductors~\cite{Sonier, Morenzoni}. Investigations of the intermediate state by $\mu$SR have been limited to a few  elemental superconductors \cite{Gladisch,Egorov,Grebinnik,Khasanov2019}; however, recently, there has been increased attention in conjunction with the nature of the superconductivity of the noncentrosymmetric BeAu \cite{Singh,Beare}. 
Here, we employed $\mu$SR spectroscopy on single crystals of ZrB$_{12}$, mapped the inhomogeneous magnetic field distributions, and identified and characterized the different superconducting states in the $H$-$T$ phase diagram. 
Our studies reveal that ZrB$_{12}$ exhibits the  coexistence of $\mu$SR signals typical of type-I and type-II superconductors. This is experimental evidence of coexisting type-I and type-II features. We theoretically reproduce this signal in a phenomenological model of a superconductor with two weakly coupled bands.

\section{Results}

In this section we present the magnetic field distribution of ZrB$_{12}$ obtained by $\mu$SR and point out the emergence of the different superconducting phases on varying the applied magnetic field and temperature. In addition, we show the direct signature of the well ordered vortex lattice appearing in the mixed state using neutron diffraction studies. This highlights the advantage of using neutrons and muons in combination to reveal different superconducting phases in materials.

\subsection{$\mu$SR studies}

Figures~\ref{fig:FFT}~\textbf{a}-\textbf{c} show the magnetic field distribution, extracted from the MaxEnt analysis of the raw TF-$\mu$SR time spectra, which are very characteristic of  the typical \textbf{a}  Meissner, \textbf{b} intermediate and \textbf{c} mixed states. At 3.6~K, 50~Oe, which is well below  $H_{\rm c1}$ at this temperature, ZrB$_{12}$ is in the Meissner state. This is clearly reflected in the observed field distribution showing a strong component at zero magnetic field. The presence of Zr nuclear moments leads to a typical slow relaxation of a Kubo-Toyabe type \cite{Kubo} for the zero-field component. 
The weak contribution at the applied field value is a background signal that is mainly due to the muons stopping in the cryostat walls and other parts of the sample holder. The absence of any additional peaks implies that the magnetic field is completely expelled from the body of the superconductor. In a type-I superconductor, demagnetisation effects may induce the intermediate state, a stable coexistence of the magnetic fields with regions of zero field and regions of internal field~${\approx\!B_{\rm c}}$. The magnetic field distribution observed at 5.2~K and 50~Oe (see Fig.~\ref{fig:FFT}~ \textbf{b}) is very similar to an intermediate state of a type-I superconductor (interestingly,  the study \cite{Junyi,Junyi2} reported vortices at these temperatures). Besides the background peak at 50~Oe, we also observe a peak at higher field $\approx\!60$~Oe, as well as a peak at ${B=0}$, indicating the presence of areas of expelled magnetic flux. At 1.6 K and an applied field of 350~Oe, ZrB$_{12}$ is found in the mixed state. Such a state is characterised by the presence of a lattice of quantized flux lines and is the hallmark of type-II superconductivity. This leads to a distribution of internal fields starting at a minimum value and increasing in weight up to the so-called saddle point of the field distribution, which is the most probable field, before falling with a long tail to a maximum field value corresponding to the region close to the vortex core. This signal is well described by a Gaussian distribution of fields centred at the saddle point, which is around 230~Oe in this case, (see Fig.~\ref{fig:FFT}~\textbf{c}), and displays the expected diamagnetic shift with respect to the applied field. We also observe a background signal at the applied field of 350~Oe. It is important to note that the absence of the peak at $B=0$ shows that the full volume of the sample is in the mixed state. 
Figure~\ref{fig:FFT}~\textbf{d} displays inhomogeneous field distribution in the phase diagram, corresponding to an unusual coexistence of the Meissner and the mixed state. This feature arises when vortices have weak attractive interaction. 
The most interesting feature is, however, presented in Fig.~\ref{fig:FFT}~\textbf{e}, dubbed as Mixed-intermediate state. This represents the coexistence of type-I and type-II superconductivity in this material. Indeed, this is experimental evidence of unique coexistent  type-I and type-II  superconducting $\mu$SR response  that manifests itself as coexistence of magnetic field excluded area as well as the two features below and above the peak corresponding to applied magnetic field. 
Figure~\ref{fig:FFT}~\textbf{f} shows that at 1.4~K, for an applied field of 500~Oe, ZrB$_{12}$ returns in the normal state. Here, the field penetrates the bulk of the sample completely, and we observe a homogeneous field distribution in the TF-$\mu$SR time spectra, corresponding to a single peak at the applied field position in the MaxEnt data~\cite{footnoteA}.

\subsection{Neutron diffraction studies}

Small-angle neutron scattering (SANS) measurements allow us to confirm and characterise the flux-line lattice (FLL) in the mixed state of ZrB$_{12}$.  A representative diffraction image of ZrB$_{12}$ is presented in Fig.~\ref{fig:FFT}~\textbf{g}. It is interesting to note that the flux-line lattice is square for all fields (applied along the \textit{a} axis) and temperatures in the mixed state of ZrB$_{12}$. In general, vortex lattices are hexagonal. Nonhexagonal vortex lattices may appear due to unconventional superconductivity~\cite{Riseman}, nonlocality~\cite{Yethiraj, Paul} and multiband effects~\cite{meng,winyard1,winyard2}. 

\section{Discussion}

\begin{figure}[b]
\begin{center}
\includegraphics[width=1.0\columnwidth]{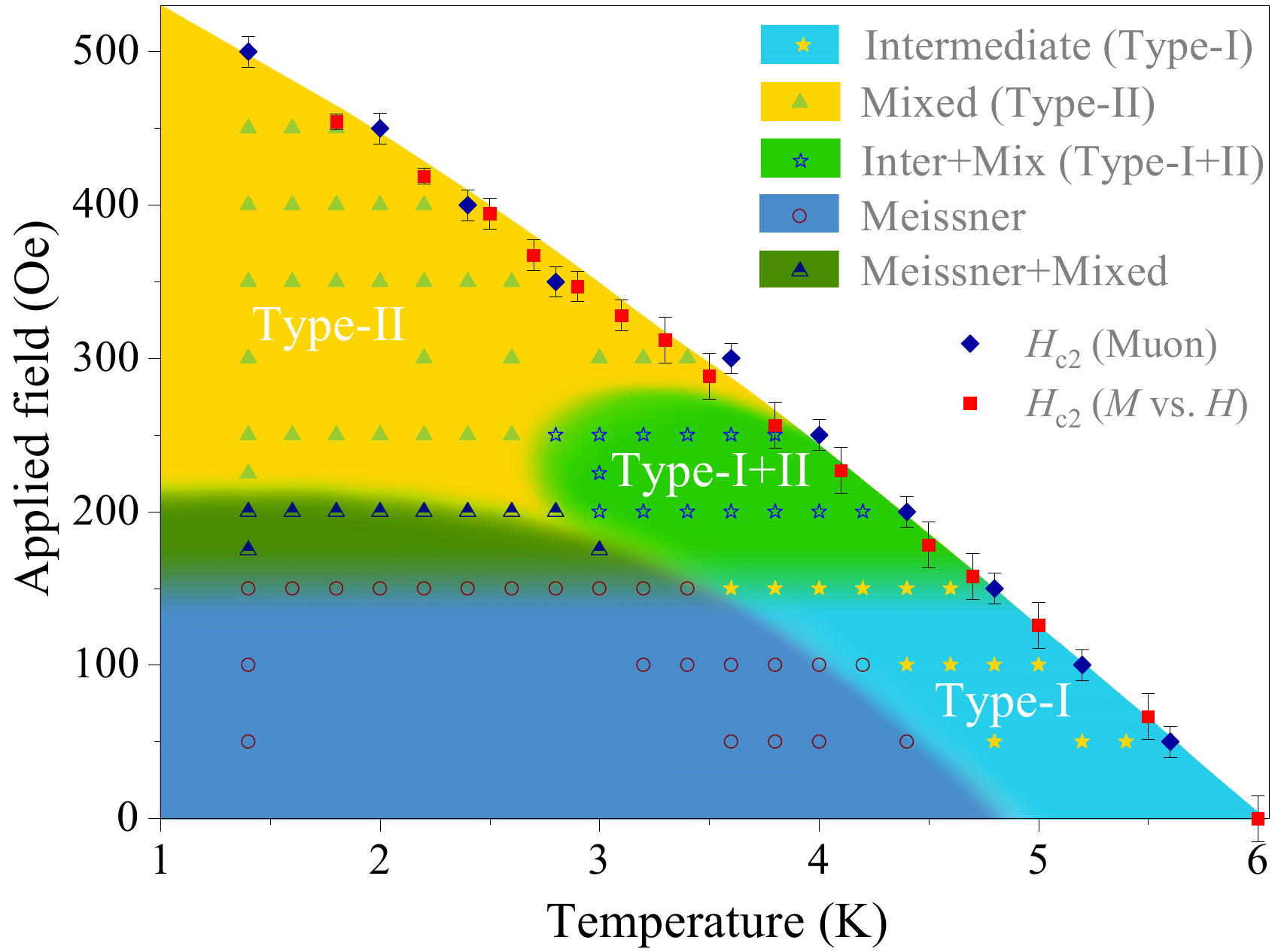}
\caption{\label{fig:phase_diagram} Different superconducting phases of ZrB$_{12}$ as described in the text; the types refer to different $\mu$SR responses shown in Fig.~\ref{fig:FFT}. Different colors approximately highlight areas with different types of $\mu$SR response. The points in the $H$-$T$ plane indicate where the data were obtained and their colors show the assignment drawn from the MaxEnt analysis of the internal field distribution.}
\end{center}
\end{figure}

TF-$\mu$SR data collected at various fields and temperature are summarized in a $H$-$T$ phase diagram in Fig.~\ref{fig:phase_diagram}. A distinctive feature of the phase diagram is the existence of three superconducting phases, giving distinctly different $\mu$SR signatures. Besides standard signatures for the Meissner, mixed, and intermediate states, we obtain unconventional response of a coexisting mixed-Meissner state. The coexistence of the vortex clusters and Meissner domains in ZrB$_{12}$ is very significant even when there is a substantial difference between the first and second critical magnetic fields. Approximating the model by a single-component theory, one can estimate $\kappa$ through the difference between the first and second critical fields. However, the width  of the parameter space where a substantial attraction is observed in single-band theory is not consistent with the single component theory of superconductors with $\kappa\approx 1/\sqrt{2}$ \cite{Eilenberger69,Jacobs73,Krageloh,Klein}. The stronger and arbitrary broad regime of intervortex attraction was predicted to occur in a very different, so-called type-1.5 state, that   forms in multiband superconductors if there are two coherence lengths one smaller and one larger than the magnetic field penetration depth \cite{Babaev2,Babaev1,Silaev1,Silaev2,PhysicaC,winyard1}. Observation of such a state was claimed for MgB$_2$ \cite{Moshchalkov, Nishio}, on the basis of surface probes showing regions with an inhomogeneous clustering of vortices and regions void of vortices. Note that the interband coupling in ZrB$_{12}$ is expected to be very weak \cite{Gasparov}. The magnetic response of such system may approximately be viewed as coexistence of almost type-I-like transport in one band (i.e., superflow that is mostly located at the surfaces of vortex clusters) and almost type-II-like transport in other band (mimicking vortices in type-II superconductors) ~\cite{Babaev_book, PhysicaC}.

The most important result of this study is the observation of the mixed-intermediate state showing coexistence of type-I-like and type-II-like responses. Note that the ``coexistent type-I and type-II response" is not a consequence of spatial inhomogeneity of $\kappa$.  Namely, in order to produce such a response, the homogeneity should be of very large length scale, with macroscopic type-I domain. But then, at different temperatures such inhomogeneity would result in multiple peaks above the applied magnetic field in the type-I regime, which was not observed here. That points to intrinsic origin of the effect.

\subsection{Theoretical modeling}

In order to analyze the origin of this $\mu$SR response, we performed numerical modeling of both single- and multiband models.
The relatively weak interband coupling, suggested by temperature dependence of magnetic field penetration length, suggests that one needs to account for superconducting degrees of freedom in both bands and use the two-band Ginzburg-Landau model~\cite{Silaev2}.
We  were able to reproduce such signatures using a phenomenological two-band model with relatively weak interband coupling in a type-1.5 regime.
The two-band Ginzburg-Landau free energy in dimensionless units  reads
\begin{align}
\mathcal{F}& = \frac{1}{2}\!\sum_{i=1,2}\big| \left(\nabla + i e {\mathbf A} \right)\psi_i\big|^2\!+\!\frac{1}{2}|\nabla\!\times\!{\mathbf A} - {\mathbf H}|^2 
- |\psi_1|^2\nonumber\\
&+ \frac{1}{2}|\psi_1|^4 + \alpha |\psi_2|^2 + \frac{\beta}{2}|\psi_2|^4 + \frac{\eta}{2}(\psi_1\psi_2^{*} + \psi_1^{*}\psi_2),
\label{GL2_dimensionless}
\end{align}
where ${\bf H}$ is applied magnetic field, ${\bf A}$ is magnetic vector potential, $\psi_{1,2}$ represent the superconducting components associated with two bands, and the interband Josephson coupling $\eta$ should be small, consistently with Ref.~\cite{Gasparov}. 
Note that a comparative study with two-band Eilenberger theory shows that this model works even at temperature far below $T_c$ if interband coupling is weak and coefficients are treated phenomenologically~\cite{Silaev2}. 
For a detailed discussion of coherence lengths in this model see, Ref.~\cite{coherencelength}.
The dimensionless parameters were searched phenomenologically to reproduce the $\mu$SR data. 
We show the modeling reproducing the $\mu$SR signal of coexistence Meissner and vortex states and the mixed state in the Supplemental Material~\cite{suppl}. 
The field distribution qualitatively corresponding to the most unusual regime, shown in Fig.~\ref{fig:FFT}~\textbf{e}, was reproduced when the model was in the regime where one coherence length is smaller and another is larger than the magnetic field penetration length~\cite{Babaev_book,PhysicaC}, ${e\!=\!1.45}$, ${\alpha\!=\!-0.5}$, ${\beta\!=\!1}$, ${\eta\!=\!\pm0.05}$, where vortices have long-range attractive and short-range repulsive interaction.
Note that for strong intervortex attraction the model has a magnetic response very similar to type-I superconductors, where instead of normal domains one has tightly bound vortex clusters.
To search for stable states, we numerically minimized the total energy, ${E\!=\!\int\!\mathcal{F}dxdy}$.
Minimization was performed by the nonlinear conjugate gradient method implemented for the NVIDIA CUDA architecture; see Ref.~\cite{Excalibur} and the Supplemental Material~\cite{suppl} for details.

\begin{figure}[ht]
\begin{center}
\includegraphics[width=1.0\columnwidth]{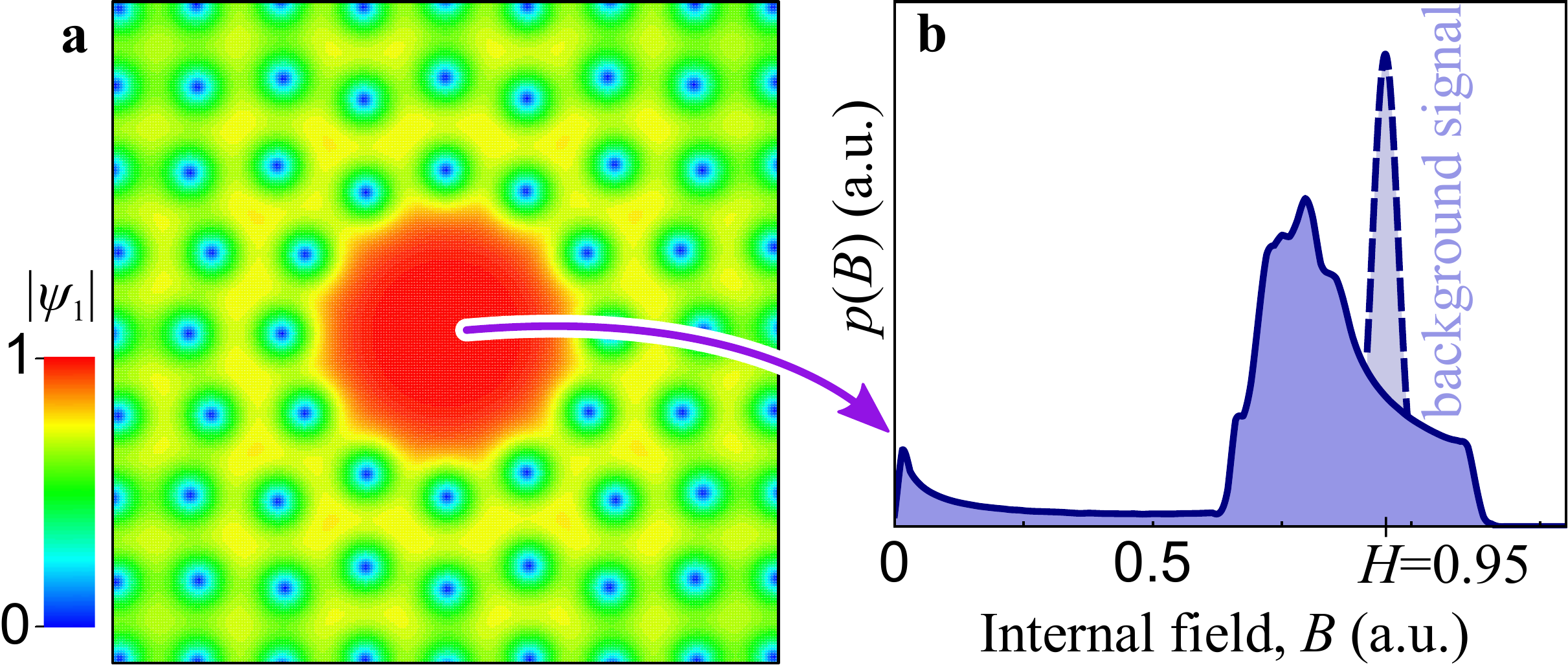}
\caption{\label{fig:simulations} 
Numerical simulation reproducing $\mu$SR signature of the mixed-intermediate state in the type-1.5 regime. \textbf{a} The state corresponding to such a response is Meissner voids in tightly bound vortex domain (a vortex analogies of Landau state). The image shows a part of the  simulation box that contained 72 vortices and was stabilized within a square whose area is approximately 1.5 times larger than the image. \textbf{b} Internal field distribution, corresponding to the area depicted in panel~\textbf{a}.}
\end{center}
\end{figure}

We find that in the type-1.5 regime there are stable states that give field distribution qualitatively similar to that observed in experiment.
Figure~\ref{fig:simulations} shows an example of a stable state that gives a similar $\mu$SR response, termed the mixed-intermediate state. 
In our simulation, we obtain such s response where a tightly bound vortex state coexistd with vortex less domains forming due to intervortex interaction. The flux exclusion domains correspond to low- and zero-magnetic field contributiond to $p(B)$. 
The tightly bound vortex state where the type-I-like component has excess current on the surface of vortex clusters give both the mixed-state-like feature $p(B)$ and the contribution in $p(B)$ corresponding to fields larger than the applied magnetic field~\cite{footnoteB}.

\section{conclusion}

We have performed $\mu$SR and neutron diffraction measurements to study the rich superconducting phase diagram of ZrB$_{12}$. The most striking  effect found in ZrB$_{12}$, is a simultaneous demonstration of both type-I and type-II characteristics in the $\mu$SR response in the same compound in a certain temperature range. It was possible to reproduce such a signal in a phenomenological model that takes into account multiple bands. At a large length scale, the state will have great similarities with the Landau state of type-I superconductors with Meissner domains immersed into proliferated tightly bound vortex domains. It would be interesting to extend our studies using low-energy-muon beams~\cite{Morenzoni} to probe the near-surface region in order to map and compare the superconductivity in this region with the bulk and to address the role of multi band effects on surface superconductivity.

\section{Acknowledgments}

The $\mu$SR experiments were performed at the ISIS Pulsed Neutron and Muon Source~\cite{isis}, STFC Rutherford Appleton Laboratory, Didcot, United Kingdom. The small angle neutron diffraction experiments were performed at the Institut Laue Langevin, Grenoble, France. This work was supported by the Engineering and Physical Sciences Research Council (EPSRC) at Warwick through Grant No. EP/I007210/1 and the Science and Technology Facilities Council (STFC) of the UK. P.K.B. would like to thank the Midlands Physics Alliance Graduate School (MPAGS) for financial support. We also wish to acknowledge useful comments from Vladimir Kogan of Ames Laboratory. S.M. acknowledges financial support from the European Unions Horizon 2020 research and innovation programme under the Marie Skodowska-Curie Grant Agreement (GA) No. 665593 awarded to the Science and Technology Facilities Council.
E.B. and F.N.R. were supported by the Swedish Research Council Grants No. 642-2013-7837, 2016-06122, 2018-03659, and G\"{o}ran Gustafsson Foundation for Research in Natural Sciences and Medicine and Olle Engkvists Stiftelse.


\begin{thebibliography}{99}
\bibitem{footnoteA}
\ITD{We want to point out that in addition to the unique ability of MaxEnt analysis to analyze the $\mu$SR spectra, it has certain limitations. 
%
For example, in Fig.~\ref{fig:FFT} it somewhat underestimates the amplitudes of low-field. 
%
This will not modify the claim of coexisting type-I and type-II signatures, but it is important to realize that similar observation in analysis using MaxEnt cannot be avoided.}
\bibitem{footnoteB}
\ITC{
We emphasize that a better comparison can be made by a direct simulation $\mu$SR time spectra  with inclusion of statistical fluctuation that could allow to extract significant information of the nature of multicomponent order parameter. We plan on implementing this theory extension in forthcoming simulations.}
\end{thebibliography}
\end{document}